# From the Fire: A Deeper Look at the Phoenix Stream

K. Tavangar[1,2], P. Ferguson[3,4], N. Shipp[1,2,5], A. Drlica-Wagner[1,2,6], S. Koposov[7,8,9], D. Erkal[10], E. Balbinot[11], J. García-Bellido[12], K. Kuehn[13,14], G. F. Lewis[15], T. S. Li[16,17,18,55], S. Mau[19,20], A. B. Pace[21], A. H. Riley[4], T. M. C. Abbott[22], M. Aguena[23], S. Allam[6], F. Andrade-Oliveira[23,24], J. Annis[6], E. Bertin[25,26], D. Brooks[27], D. L. Burke[20,28], A. Carnero Rosell[23], M. Carrasco Kind[29,30], J. Carretero[31], M. Costanzi[32,33,34], L. N. da Costa[23,35], M. E. S. Pereira[36,37], J. De Vicente[38], H. T. Diehl[6], S. Everett[39], I. Ferrero[40], B. Flaugher[6], J. Frieman[2,6], E. Gaztanaga[41,42], D. W. Gerdes[36,43], D. Gruen[44], R. A. Gruendl[29,30], J. Gschwend[23,35], G. Gutierrez[6], S. R. Hinton[45], D. L. Hollowood[39], K. Honscheid[46,47], D. J. James[48], N. Kuropatkin[6], M. A. G. Maia[23,35], J. L. Marshall[4], F. Menanteau[29,30], R. Miquel[31,49], R. Morgan[3], R. L. C. Ogando[35], A. Palmese[2,6], F. Paz-Chinchón[8,29], A. Pieres[23,35], A. A. Plazas Malagón[50], M. Rodriguez-Monroy[38], E. Sanchez[38], V. Scarpine[6], S. Serrano[41,42], I. Sevilla-Noarbe[38], M. Smith[51], E. Suchyta[52], M. E. C. Swanson[29], G. Tarle[36], C. To[19,20,28], T. N. Varga[53,54], and A. R. Walker[22]

(DES Collaboration)

[1] Department of Astronomy and Astrophysics, University of Chicago, Chicago, IL 60637, USA; ktavangar@flatironinstitute.org
[2] Kavli Institute for Cosmological Physics, University of Chicago, Chicago, IL 60637, USA
[3] Physics Department, 2320 Chamberlin Hall, University of Wisconsin−Madison, 1150 University Avenue Madison, WI 53706-1390, USA
[4] George P. and Cynthia Woods Mitchell Institute for Fundamental Physics and Astronomy, and Department of Physics and Astronomy, Texas A&M University, College Station, TX 77843, USA
[5] MIT Kavli Institute for Astrophysics and Space Research, 77 Massachusetts Ave., Cambridge, MA 02139, USA
[6] Fermi National Accelerator Laboratory, P.O. Box 500, Batavia, IL 60510, USA
[7] Institute for Astronomy, University of Edinburgh, Royal Observatory, Blackford Hill, Edinburgh, EH9 3HJ, UK
[8] Institute of Astronomy, University of Cambridge, Madingley Road, Cambridge, CB3 0HA, UK
[9] Kavli Institute for Cosmology, University of Cambridge, Madingley Road, Cambridge, CB3 0HA, UK
[10] Department of Physics, University of Surrey, Guildford, GU2 7XH, UK
[11] Kapteyn Astronomical Institute, University of Groningen, Postbus 800, NL-9700AV Groningen, The Netherlands
[12] Instituto de Fisica Teorica UAM/CSIC, Universidad Autonoma de Madrid, E-28049 Madrid, Spain
[13] Lowell Observatory, 1400 Mars Hill Rd., Flagstaff, AZ 86001, USA
[14] Australian Astronomical Optics, Macquarie University, North Ryde, NSW 2113, Australia
[15] Sydney Institute for Astronomy, School of Physics, A28, The University of Sydney, NSW 2006, Australia
[16] Department of Astronomy and Astrophysics, University of Toronto, 50 St. George Street, Toronto, ON M5S 3H4, Canada
[17] Observatories of the Carnegie Institution for Science, 813 Santa Barbara St., Pasadena, CA 91101, USA
[18] Department of Astrophysical Sciences, Princeton University, Princeton, NJ 08544, USA
[19] Department of Physics, Stanford University, 382 Via Pueblo Mall, Stanford, CA 94305, USA
[20] Kavli Institute for Particle Astrophysics & Cosmology, P.O. Box 2450, Stanford University, Stanford, CA 94305, USA
[21] Department of Physics, Carnegie Mellon University, Pittsburgh, PA 15312, USA
[22] Cerro Tololo Inter-American Observatory, NSF's National Optical-Infrared Astronomy Research Laboratory, Casilla 603, La Serena, Chile
[23] Laboratório Interinstitucional de e-Astronomia—LIneA, Rua Gal. José Cristino 77, Rio de Janeiro, RJ—20921-400, Brazil
[24] Instituto de Física Teórica, Universidade Estadual Paulista, São Paulo, Brazil
[25] CNRS, UMR 7095, Institut d'Astrophysique de Paris, F-75014, Paris, France
[26] Sorbonne Universités, UPMC Univ Paris 06, UMR 7095, Institut d'Astrophysique de Paris, F-75014, Paris, France
[27] Department of Physics & Astronomy, University College London, Gower Street, London, WC1E 6BT, UK
[28] SLAC National Accelerator Laboratory, Menlo Park, CA 94025, USA
[29] Center for Astrophysical Surveys, National Center for Supercomputing Applications, 1205 West Clark St., Urbana, IL 61801, USA
[30] Department of Astronomy, University of Illinois at Urbana-Champaign, 1002 W. Green Street, Urbana, IL 61801, USA
[31] Institut de Física d'Altes Energies (IFAE), The Barcelona Institute of Science and Technology, Campus UAB, E-08193 Bellaterra (Barcelona) Spain
[32] Astronomy Unit, Department of Physics, University of Trieste, via Tiepolo 11, I-34131 Trieste, Italy
[33] INAF-Osservatorio Astronomico di Trieste, via G.B. Tiepolo 11, I-34143 Trieste, Italy
[34] Institute for Fundamental Physics of the Universe, Via Beirut 2, I-34014 Trieste, Italy
[35] Observatório Nacional, Rua Gal. José Cristino 77, Rio de Janeiro, RJ—20921-400, Brazil
[36] Department of Physics, University of Michigan, Ann Arbor, MI 48109, USA
[37] Hamburger Sternwarte, Universität Hamburg, Gojenbergsweg 112, D-21029 Hamburg, Germany
[38] Centro de Investigaciones Energéticas, Medioambientales y Tecnológicas (CIEMAT), Madrid, Spain
[39] Santa Cruz Institute for Particle Physics, Santa Cruz, CA 95064, USA
[40] Institute of Theoretical Astrophysics, University of Oslo, P.O. Box 1029 Blindern, NO-0315 Oslo, Norway
[41] Institut d'Estudis Espacials de Catalunya (IEEC), E-08034 Barcelona, Spain
[42] Institute of Space Sciences (ICE, CSIC), Campus UAB, Carrer de Can Magrans, s/n, E-08193 Barcelona, Spain
[43] Department of Astronomy, University of Michigan, Ann Arbor, MI 48109, USA
[44] Faculty of Physics, Ludwig-Maximilians-Universität, Scheinerstr. 1, D-81679 Munich, Germany
[45] School of Mathematics and Physics, University of Queensland, Brisbane, QLD 4072, Australia
[46] Center for Cosmology and Astro-Particle Physics, The Ohio State University, Columbus, OH 43210, USA
[47] Department of Physics, The Ohio State University, Columbus, OH 43210, USA
[48] Center for Astrophysics | Harvard & Smithsonian, 60 Garden Street, Cambridge, MA 02138, USA
[49] Institució Catalana de Recerca i Estudis Avançats, E-08010 Barcelona, Spain

[50] Department of Astrophysical Sciences, Princeton University, Peyton Hall, Princeton, NJ 08544, USA
[51] School of Physics and Astronomy, University of Southampton, Southampton, SO17 1BJ, UK
[52] Computer Science and Mathematics Division, Oak Ridge National Laboratory, Oak Ridge, TN 37831, USA
[53] Max Planck Institute for Extraterrestrial Physics, Giessenbachstrasse, D-85748 Garching, Germany
[54] Universitäts-Sternwarte, Fakultät für Physik, Ludwig-Maximilians Universität München, Scheinerstr. 1, D-81679 München, Germany

## Abstract

We use 6 yr of data from the Dark Energy Survey to perform a detailed photometric characterization of the Phoenix stellar stream, a $15°$ long, thin, dynamically cold, low-metallicity stellar system in the Southern Hemisphere. We use natural splines, a nonparametric modeling technique, to simultaneously fit the stream track, width, and linear density. This updated stream model allows us to improve measurements of the heliocentric distance ($17.4 \pm 0.1$ (stat.) $\pm$ 0.8 (sys.) kpc) and distance gradient ($-0.009 \pm 0.006$ kpc deg$^{-1}$) of Phoenix, which corresponds to a small change of $0.13 \pm 0.09$ kpc in heliocentric distance along the length of the stream. We measure linear intensity variations on degree scales, as well as deviations in the stream track on $\sim 2°$ scales, suggesting that the stream may have been disturbed during its formation and/or evolution. We recover three peaks and one gap in linear intensity along with fluctuations in the stream track. Compared to other thin streams, the Phoenix stream shows more fluctuations and, consequently, the study of Phoenix offers a unique perspective on gravitational perturbations of stellar streams. We discuss possible sources of perturbations to Phoenix, including baryonic structures in the Galaxy and dark matter subhalos.

*Key words:* Cosmology – Dark matter – Stellar streams – Galaxy structure – Astronomy data modeling – Milky Way dynamics

## 1. Introduction

Near-field cosmology utilizes observations of small-scale cosmological structures to answer fundamental questions about the composition and evolution of our universe. One of the major goals of near-field cosmology is to measure the distribution of dark matter in the local universe, in order to improve our understanding of galaxy formation, structure, and evolution (Conroy & Wechsler 2009; Guo et al. 2013; Becker 2015; Wechsler & Tinker 2018). Stellar streams, the tidally disrupted remnants of satellite galaxies and globular clusters, provide an exciting avenue in the study of near-field cosmology (e.g., Bullock & Johnston 2005; Carlberg 2012; Bovy et al. 2016; Malhan & Ibata 2019). Their abundance and orbital histories enable tests of galaxy formation, accretion history, and stellar halo formation (e.g., Johnston 1998; Helmi & White 1999; Bonaca et al. 2019; Malhan et al. 2021). Dynamically cold streams originating from disrupting star clusters are extremely sensitive to gravitational perturbations from massive structures (e.g., Johnston et al. 2002; Erkal et al. 2019; Shipp et al. 2021; Vasiliev et al. 2021) and substructures (e.g., Ibata et al. 2002; Erkal & Belokurov 2015a; Bonaca et al. 2019). The study of stellar streams is thus a promising avenue for studying the distribution of dark matter at subgalactic scales.

Large additional sky surveys have advanced rapidly since the Sloan Digital Sky Survey was used to discover tidal tails emanating from the Palomar 5 globular cluster (Odenkirchen et al. 2001, 2002, 2003; Rockosi et al. 2002). The known population of stellar streams has drastically expanded in recent years owing to the development of new data sets and analysis techniques. The increase in survey coverage and sensitivity has allowed us to probe a larger volume of the Milky Way halo and a wider range of the stream parameter space

(Belokurov et al. 2006; Bernard et al. 2016; Malhan & Ibata 2018; Ibata et al. 2019, 2021). In particular, deep imaging and precise photometry from the Dark Energy Survey (DES; DES Collaboration 2005, 2016) have yielded a large and diverse population of stellar streams in the Southern Hemisphere (Drlica-Wagner et al. 2015; Balbinot et al. 2016; Shipp et al. 2018).

Among the streams discovered by DES, the Phoenix stream is of particular interest owing to its prominence, clumpy morphology, and low metallicity. Phoenix is kinematically cold, with a velocity dispersion of $\sigma_{RV} = 2.66$ km s$^{-1}$ (Wan et al. 2020). The $15°$ ($\sim$4.6 kpc) long stream was discovered in early DES data by Balbinot et al. (2016). Broadly speaking, streams originating from globular clusters (e.g., Palomar 5) are thinner than those generated from dwarf galaxies (e.g., Sagittarius; see Grillmair & Carlin 2016, pp. 87–114). The narrow width of Phoenix ($\sim$0°.14, $\sim$43 pc) suggests that it likely originated from a disrupted globular cluster. The globular cluster progenitor hypothesis is supported by stellar metallicity measurements by Wan et al. (2020), who found that the metallicity spread for stars in Phoenix was $\sigma_{[Fe/H]} \approx 0$. If Phoenix originated from a globular cluster, then it would be among the most metal-poor globular clusters known, with a metallicity of $[Fe/H] = -2.7$ dex. However, no clear progenitor for the Phoenix stream has been found (Balbinot et al. 2016).

Regardless of their origin, cold, thin streams like Phoenix are excellent structures for probing the distribution of dark matter in the Milky Way halo. Detailed analyses of the spatial structure of stellar streams can give valuable information about the abundance of dark matter subhalos, which are predicted to be plentiful in the conventional model of cold dark matter (CDM; Klypin et al. 1999; Diemand et al. 2005; Montanari & García-Bellido 2020; Wang et al. 2020). Measurements of streams may be able to detect the statistical influence of subhalos on the disruption of baryonic structures (e.g., Johnston et al. 2002; Carlberg 2012; Erkal & Belokurov 2015b; Banik et al. 2018) and also the density and localization of individual perturbers within the Galactic halo (e.g., Erkal & Belokurov 2015b; Bonaca et al. 2019). To date, a relatively small number of cold, thin streams have had detailed morphological measurements. This list currently includes Palomar 5 (Odenkirchen et al. 2001; Erkal et al. 2017), Orphan (Grillmair 2006; Koposov et al. 2019), ATLAS and Aliqa Uma (Koposov et al. 2014; Li et al. 2020), and Jet (Jethwa et al. 2018; Ferguson et al. 2021). We note that the GD-1 stream has also been modeled, although with a different method compared to the other streams (de Boer et al. 2018). Expanding the set of consistently characterized streams will allow us to compare density fluctuations and wiggles across streams, to test different perturber models, and to learn more about the frequency of stream interactions in our Milky Way.

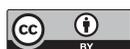

In this paper, we use a photometric catalog derived from 6 yr of DES imaging to characterize the Phoenix stream in a manner consistent with the analyses performed on other streams. We present and model detailed observational data, obtaining measurements of the stream track, width, intensity, and distance gradient. We also model the orbit of the stream in the Milky Way potential using member star radial velocities and proper motions from the Southern Stellar Stream Spectroscopic Survey ($S^5$; Li et al. 2019) and Gaia EDR3 (Gaia Collaboration et al. 2016; Gaia Collaboration et al. 2021), respectively.

The structure of this paper is as follows. In Section 2, we describe the DES data used in this analysis. In Section 3, we describe the process used to generate an optimized matched filter for selecting stream members. In Section 4, we present our model for the stream track, width, and intensity. We then apply that model to measure the distance gradient of Phoenix and fit a 6D orbit model. In Section 5, we discuss the density and stream track variations detected in the models. Finally, we conclude in Section 6.

## 2. Data Set

DES is a 6 yr optical/near-infrared imaging survey covering $\sim$5000 deg$^2$ of the southern Galactic cap in five visible/near-infrared filters, $grizY$, using the Dark Energy Camera (DECam; Flaugher et al. 2015) mounted at the prime focus of the 4 m Blanco telescope at the Cerro Tololo Inter-American Observatory. The uniform, wide-area imaging of DES allows for deep searches for stellar streams over a large section of the southern sky and has resulted in numerous stream discoveries (e.g., Shipp et al. 2018).

In this paper, we use catalog data derived from the full 6 yr DES co-added images (DES Collaboration et al. 2021). The DES images are processed using the DES data management pipeline (Morganson et al. 2018). After photometric calibration using the Forward Global Calibration Method (FGCM; Burke et al. 2017), the images are co-added to increase imaging depth. Our data set follows the same image-level processing and co-addition described in DES Collaboration et al. (2021); however, it is augmented with value-added properties derived from multiepoch fitting used for DES cosmology analyses (e.g., Drlica-Wagner et al. 2018; Sevilla-Noarbe et al. 2020). This pipeline performs a simultaneous fit to the individual images using their respective point-spread functions and thus provides improved point-source photometry relative to fits performed on the co-added images. We apply a morphological filter to select high-probability stellar sources by using an updated version of the `EXTENDED_CLASS` classifier defined in the DES Y3 Gold release, as described in Section 6.1 and Appendix B of Sevilla-Noarbe et al. (2020). The classifier is adapted to incorporate updates to the DES simultaneous multiepoch photometry, known as the single-object fit (SOF) pipeline, applied to the deeper Y6 data (see Section 3.3 of Sevilla-Noarbe et al. 2020 for a description of the SOF pipeline). This is achieved in practice by taking `EXT_SOF < 2` to select a relatively complete sample of stars.

Our DES data set contains five additional years of observations relative to the initial analysis of the Phoenix stream in Balbinot et al. (2016) and three additional years of observations relative to the analysis Shipp et al. (2018). The signal-to-noise ratio (S/N) = 10 magnitude limits of our stellar sample are $g = 24.4$ mag and $r = 24.2$ mag. This corresponds to S/N = 12 in the $g - r$ color of the main sequence of Phoenix at $g = 24$ mag. Our data set is $\sim$0.7 mag deeper than that of Balbinot et al. (2016), which is consistent with roughly doubling the total exposure time in this region of the sky (the first year of DES observations covered a fraction of the DES footprint to about half the final survey depth). Furthermore, the multiepoch star/galaxy classification described above increases the completeness and purity of our stellar sample relative to Balbinot et al. (2016), who used a selection based on the weighted average of the single-epoch quantities. We perform our analysis of the Phoenix stream on a stellar sample with $g < 24$ mag, which is 1 mag deeper than the selection performed by Balbinot et al. (2016). As a result, the new DES data enable increased accuracy for characterizing the Phoenix stellar stream.

## 3. Methods and Analysis

To characterize the stellar population of the Phoenix stream, we perform a matched-filter selection in color–magnitude space. In order to define this selection, we choose an initial stream track using the width and stream endpoints taken from Shipp et al. (2018). We use a stream coordinate system, ($\phi_1$, $\phi_2$), defined by the rotation matrix reported in Appendix D of Shipp et al. (2019).[56] We define an "off-stream" region of the same length and width, but offset by one degree in stream coordinates above (roughly west) Phoenix. We create a background-subtracted Hess diagram by subtracting the on- and off-stream regions (Figure 1).[57] This data-driven selection statistically removes stars from the more diffuse Eri-Phe stellar cloud, which spatially overlaps the Phoenix stream and resides at a similar distance of $\sim$16 kpc (Li et al. 2016).

Following the analysis in Section 3.1 of Shipp et al. (2018), we perform a binned maximum likelihood fit of the smoothed two-dimensional background-subtracted Hess diagram using a synthetic isochrone from Dotter et al. (2008) as implemented in `ugali` (Bechtol et al. 2015; Drlica-Wagner et al. 2020).[58] We fix the metallicity of the isochrone at the most metal-poor value provided by Dotter et al. (2008), which has a metallicity of $Z = 0.00007$ ([Fe/H] $\approx -2.5$). This is slightly more metal-rich than the extremely low spectroscopic metallicity of Phoenix ($Z = 0.00004$, [Fe/H] $= -2.7$; Wan et al. 2020); however, more metal-rich isochrones do a good job of fitting the photometry of the spectroscopically confirmed members (see Extended Data Figure 1 in Wan et al. 2020). This fixed metallicity allows us to forgo selections on the red giant branch

---

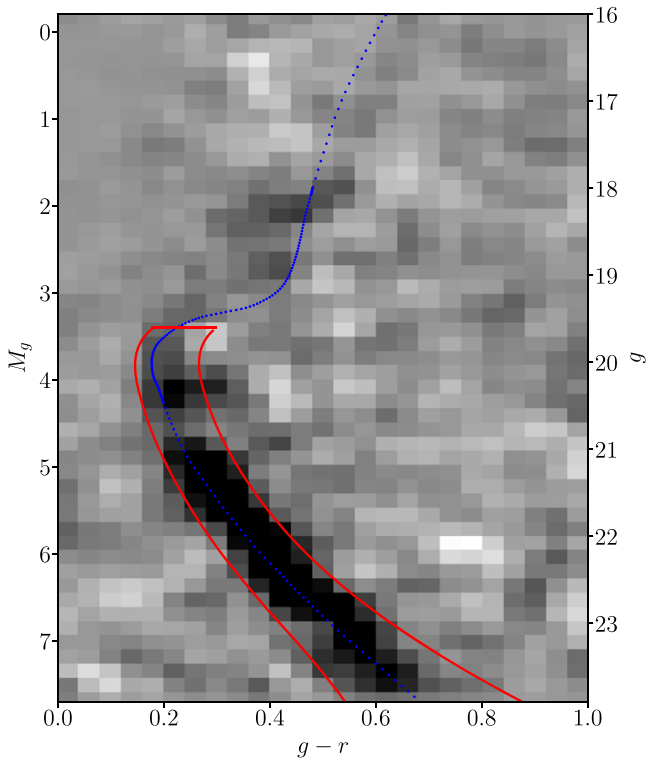

**Figure 1.** A Hess difference plot created by subtracting a background region above Phoenix (in $\phi_2$) of equal area to the stream. The main sequence is clearly visible. The Dotter et al. (2008) isochrone is shown in blue with parameters of [Fe/H] = −2.5, $\tau$ = 12.8 Gyr, and $m - M = 16.2$. This isochrone is selected as part of an iterative process where we take the output of the stream model described in Section 4 and find its distance gradient and average distance modulus, which informs an updated isochrone selection. The red outline corresponds to a filter in color–magnitude space used to select stars associated with the Phoenix stream, guided by the best-fit isochrone. We determine the parameters of this isochrone selection region as $\Delta\mu = 0.5$, $C_{1,2} = (0.01, 0.1)$, and $E = 2$ via visual inspection of the Hess diagram. The on-stream region is selected in a similar iterative way using the stream track and stream width outputs of the model.

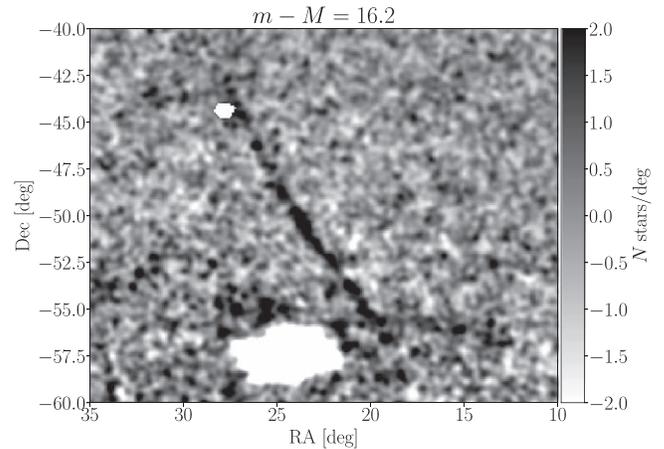

**Figure 2.** A spatial density map of the region surrounding the Phoenix stream showing the background-subtracted number of stars per pixel. The pixel size is $0°.1 \times 0°.1$, and a Gaussian filter with a smoothing kernel of $0°.15$ is used to smooth the residual stellar counts. The map is made using a matched-filter selection described in Section 3, with isochrone parameters [Fe/H] = −2.7, $\tau$ = 12.8 Gyr, and $m - M = 16.2$. We note a dark horizontal feature to the southwest of the stream at decl. $\approx -55°$. This feature resides at a similar distance modulus of $m - M = 16.2$, but it is not believed to be related to the Phoenix stream owing to its differing track orientation. Two white circles indicate masked regions around the Phoenix dwarf galaxy (R.A., decl. = $27°.26$, $-44°.69$) and the bright star Achernar (R.A., decl. = $24°.43$, $-57°.24$).

that would normally provide a better metallicity estimate but would decrease the purity of our sample of Phoenix members, which we deem more important for our models. We use an affine-invariant Markov Chain Monte Carlo (MCMC) ensemble sampler (emcee; Foreman-Mackey et al. 2013) to derive the posterior probability distributions for the age, distance modulus, and richness (number of stars) of the Phoenix stream assuming uniform priors ($\tau = [10$ Gyr, $14$ Gyr], $m - M = [14.2$ mag, $18.2$ mag], richness >0) in each parameter. This fit yields a best-fit age of $\tau = 12.8 \pm 0.2$ Gyr, a distance modulus of $m - M = 16.3 \pm 0.1$ mag, and a richness of $16,000 \pm 2000$ stars.

We select stars associated with the Phoenix stream by defining a filter in color–magnitude space guided by the best-fit isochrone. Once again following the process described in Section 3.1 of Shipp et al. (2018), we define a selection region around the synthetic isochrone based on a symmetric magnitude broadening, $\Delta\mu$, an asymmetric color broadening, $C_{1,2}$, and a multiplicative factor for broadening based on the photometric uncertainty, $E$ (see Equation (4) of Shipp et al. 2018). We determine the parameters of the isochrone selection region as $\Delta\mu = 0.5$, $C_{1,2} = (0.01, 0.1)$, and $E = 2$ via visual inspection of the Hess diagram. We also limit our selection to an absolute magnitude of $3.4 < M_g < 7.8$ to ensure that we select stars along the main sequence, which have the highest contrast relative to the background color–magnitude distribution. We note that throughout this process we do not explicitly broaden our selection to account for binary stars, but they should be included within our isochrone selection region.

To visualize the track of the Phoenix stream on the sky, we scan our isochrone filter across a range of distance moduli, from $15 < m - M < 20$ in steps of size 0.1 mag. We fit a fifth-order polynomial to the smooth background at each distance modulus and subtract its contribution from the isochrone-filtered stellar density. We smooth the residual stellar counts by a Gaussian filter with a smoothing kernel of size $0°.15$. The residual stellar density map at a distance modulus of $m - M = 16.2$ is shown in Figure 2. From a visual inspection of maps at each distance modulus, we confirm that the significance of the signal from the Phoenix stream is maximized at $m - M = 16.2$, which is in good agreement with our quantitative analysis in Section 4.2 and the value of $m - M = 16.21 \pm 0.11$ mag reported by Balbinot et al. (2016). At this distance, we find an excess of 819 stars passing our isochrone selection and located along the stream relative to the expected number based on the background polynomial fit.

We note the existence of a feature to the southwest of the stream in Figure 2 at decl. $\approx -55°$. We believe this structure to be real, and it does reside at a similar distance modulus of $m - M = 16.2$, but because of its orientation, we do not consider it to be part of the Phoenix stream. However, we do note that similar kinks have been seen recently in stellar streams such as the ATLAS Aliqa Uma system (Li et al. 2020). Photometry alone is not enough to confirm a relationship between such features, and we leave the spectroscopic follow-up of this structure to future work.

We apply our initial color–magnitude filter to select stars for an initial fit to the Phoenix stream morphology (Section 4.1) and distance gradient (Section 4.2). These improved models allow us to recover a more accurate distance modulus for the stream, which in turn lets us improve our isochrone selection. We repeat this iterative process of modifying the isochrone selection and refitting the stream until the distance modulus recovered from the distance gradient model is within the statistical uncertainty of the one used in the matched-filter selection. This iterative process converges to yield a distance modulus value of $16.20 \pm 0.01$ (stat.) $\pm 0.1$ (sys.) mag ($17.4 \pm 0.1$ (stat.) $\pm 0.8$ (sys.) kpc), which we use for the final version of our matched-filter selection shown in Figure 1. The above estimate includes a formal statistical error derived from the procedure in Section 4.2 and a systematic error derived by Drlica-Wagner et al. (2015) from fitting different synthetic isochrone models to faint stellar systems in DES.

## 4. Results

### 4.1. Stream Track Model

In order to characterize the morphology of the Phoenix stream, we build a spline model for the stream track, width, and intensity. Our procedure is based on similar models introduced in Erkal et al. (2017) and Koposov et al. (2019), and is nearly identical in setup to the model used in Li et al. (2020). Our model uses natural cubic splines with varying numbers of nodes along the stream track coordinate, $\phi_1$, to describe the data in the region surrounding the Phoenix stream. To simplify the fitting process, we assume the linear density profile of the stream to be Gaussian in $\phi_2$, such that the stream model can be described by three parameters: the logarithm of the stream width, the logarithm of the stream intensity, and the stream track. We also fit the logarithm of the background density, described by a quadratic in $\phi_2$. Thus, the full stream and background model is described by six parameters:

$$
\begin{aligned}
\rho(\phi_1, \phi_2) = {} & \exp[\beta_0(\phi_1) + \phi_2\beta_1(\phi_1) + \phi_2^2\beta_2(\phi_1)] \\
& + \exp[I(\phi_1)] \exp\left[-\frac{1}{2}\left(\frac{\phi_1 - \Phi_2(\phi_1)}{\exp[S(\phi_1)]}\right)^2\right].
\end{aligned}
\tag{1}
$$

In this equation, $\beta_0$, $\beta_1$, and $\beta_2$ represent the normalization, slope, and quadratic term of the log-background, while $\Phi_2$, $S$, and $I$ represent the stream track, the stream width, and the linear stream density, respectively. We implement this model in the STAN programming language, which is specialized for statistical modeling and provides an easy interface to specify and efficiently sample probabilistic models (Carpenter et al. 2017). The posterior distributions of the model parameters given the Phoenix data are sampled using the No-U-Turn Sampler version of the Hamilton Monte Carlo technique, which is useful for exploring high-dimensional parameter spaces efficiently (Hoffman & Gelman 2011; Neal 2011, pp. 113–162; Betancourt 2018).

The model also requires the placement of nodes for each of the six model parameters. From the model posterior, we determine the optimal value of the given parameter at each node and then construct the cubic spline. We simplify the process of node placement by constraining the nodes to be equidistant for each parameter. This allows us to define their positions only by specifying the number of nodes. We then run a Bayesian optimization to select the optimal number of nodes for each parameter using GPyOpt.[59] We verify our implementation of this optimization scheme by applying it to the

ATLAS stream and finding good agreement with the results of Li et al. (2020) (see the Appendix for details).

Applying these methods to the Phoenix stream in the DES Y6 data, we find that the stream is best described by 14, 7, 20, 15, 8, and 6 nodes for the stream track, width, linear intensity, background density, slope of the log-background, and quadratic term of the log-background, respectively.[60] We now explore the posterior probability distributions for each of the model parameters by running 12 sampling chains for 5000 iterations, discarding the first 2000 iterations as burn-in. We ensure that each chain has converged based on the Gelman–Rubin convergence diagnostic, $\hat{R} < 1.1$ (Gelman & Rubin 1992).

The best-fit stream model is shown in Figure 3. We note that the model recovers a very clumpy stream, consisting of ~1000 stars, with small-scale variations in the linear intensity and track along $\phi_1$, while the width and background remained relatively constant. There are three main peaks in the stream intensity at $\phi_1 \approx -1^\circ 5$, $3^\circ$, and $5^\circ$.

We test the modeling procedure by running the same routine on a simulated stream of uniform intensity with no $\phi_2$ offset. We inject this simulated stream into the DES data located several degrees west of Phoenix and devoid of other known stellar streams. The resulting best-fit model is consistent with the simulated stream, and it shows none of the complexity that we observe for Phoenix in the stream track, intensity, or width. This gives us confidence that variations in the foreground stellar density are not contributing significantly to the complex structure we observe for Phoenix.

In addition, our model suggests small-scale variations in the stream track, which we refer to as "wiggles." These wiggles have amplitudes that are smaller than the reported stream width but are 3–4 times larger than the spatial binning of our data set, which makes it unlikely that they are binning artifacts. We estimate the statistical significance of the wiggles by comparing the log-likelihood of our full model with the log-likelihood of a model with a flat stream track constrained to three nodes. We find that the change in the log-likelihood is $\Delta \log(\mathcal{L}) = 8.4$. Interpreting the difference in the log-likelihood in terms of a $\chi^2$ distribution with degrees of freedom equal to the difference in the number of nodes in the models (14 nodes vs. 3 nodes), we find a significance of $1.2\sigma$ for the stream wiggles.

We also consider whether the stream wiggles could be artificially introduced by our modeling procedure. Recently, de Boer et al. (2018) suggested that wiggles in $\Phi_2$ could be introduced when the model has trouble locating the exact center of the density distribution at each $\phi_1$. However, we find this to be unlikely since the uncertainties on both the width and the track are rather small. Another possibility is that changes in stream intensity might artificially affect the fit of the stream track. In Phoenix, this could contribute to wiggles in the track at $\phi_1 \sim 3^\circ - 5^\circ$, coinciding with density variations in the same region. We consider possible physical explanations for the small-scale variations in the density and track in Section 5.

### 4.2. Distance Gradient

The improved quality and depth of the DES Y6 data allow us to measure the distance and distance gradient of the Phoenix stream more accurately than was possible with the DES Y1 data analyzed by Balbinot et al. (2016). To measure the distance gradient, we scan our isochrone selection in distance modulus from $14.5 \leqslant m - M \leqslant 18.0$ in steps of 0.01 mag. At each step in distance modulus, we count the number of stars on the stream track passing the isochrone selection and create a 2D histogram of those stars as a function of distance modulus and $\phi_1$. This histogram is shown in the left panel of Figure 4.

---

[60] We note that we limit our possible number of nodes to between 5 and 30 for the stream track, density, and background; between 3 and 30 for the width; between 3 and 20 for the slope of the log-background; and between 3 and 10 for the quadratic term of the log-background.

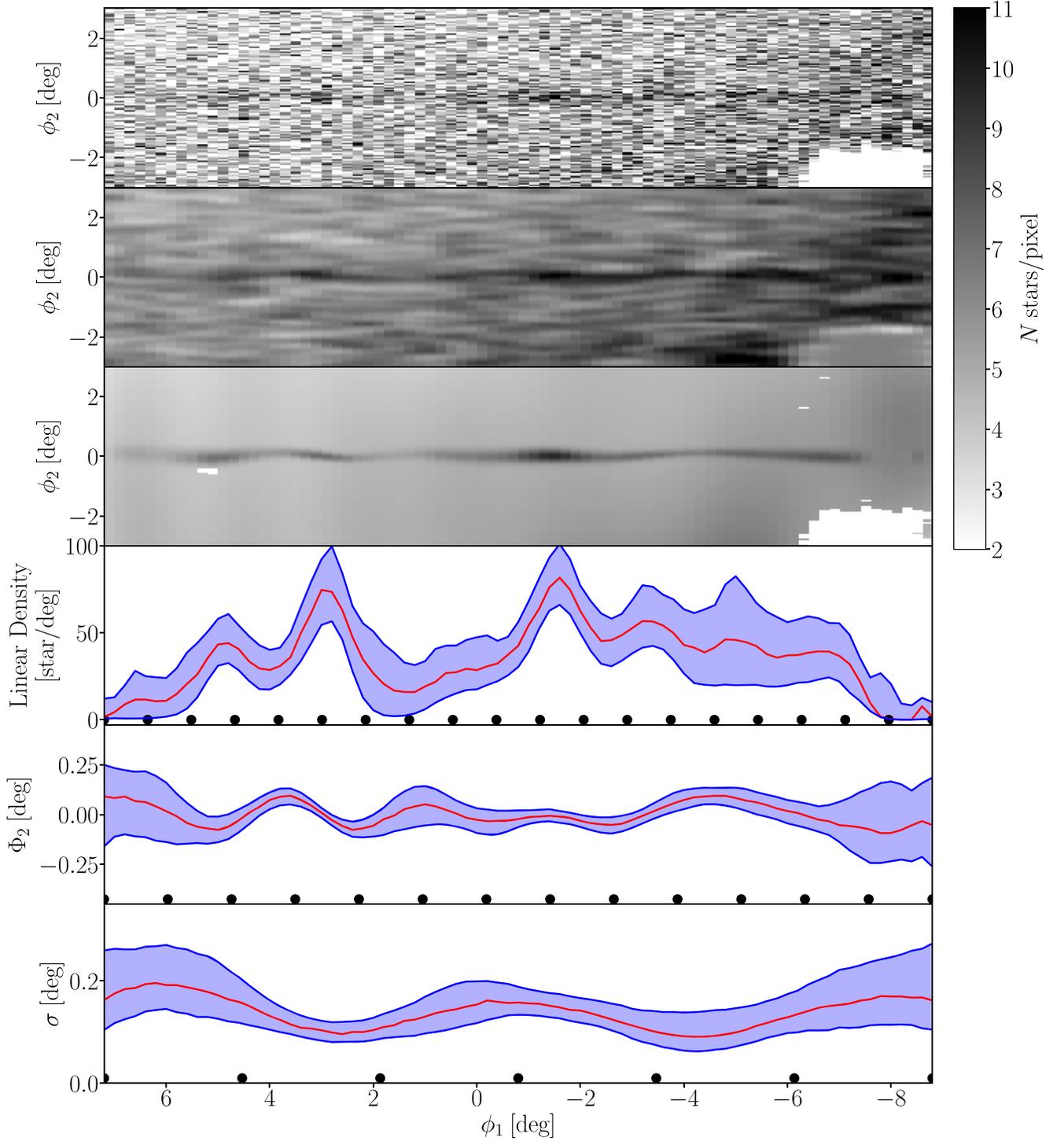

**Figure 3.** Morphological model of the Phoenix stream. Top panel: the spatial density of stars passing our isochrone matched filter binned into $0.2 \times 0.05$ deg$^2$ pixels in stream coordinates, without smoothing. Second panel: the smoothed spatial density of stars using a Gaussian filter with a smoothing kernel of $0°.15$. Third panel: the stellar density predicted by our natural cubic spline model of the stream and background (Equation (1)). Fourth panel: the stream linear density, calculated by vertically summing the stream intensity at each $\phi_1$, excluding the background. On average, the stream contains $\approx 25$ stars deg$^{-1}$ more than the off-stream background region. Prominent peaks occur at $\phi_1 \approx -1°.5$, $3°$, and $5°$. Fifth panel: the stream track as a function of $\phi_1$. The track attains maximum values of $\phi_2$ at $\phi_1 \approx -4°$, $1°$, and $3°.5$. The significance of these wiggles is $1.3\sigma$ and is discussed in Section 5.3. Bottom panel: the stream width as a function of $\phi_1$. In each of the lower three panels, the red lines represent the peak of a KDE fit to the sample outputs for each parameter. The blue shaded regions show the minimum interval containing 68% of the posterior probability. The black circles represent the locations of the nodes used to measure each parameter. In all panels the stream coordinate system is defined by the rotation matrix in Appendix D of Shipp et al. (2019), which relates to the one used in Balbinot et al. (2016) by $\phi_1 \approx 285° - \Lambda$.

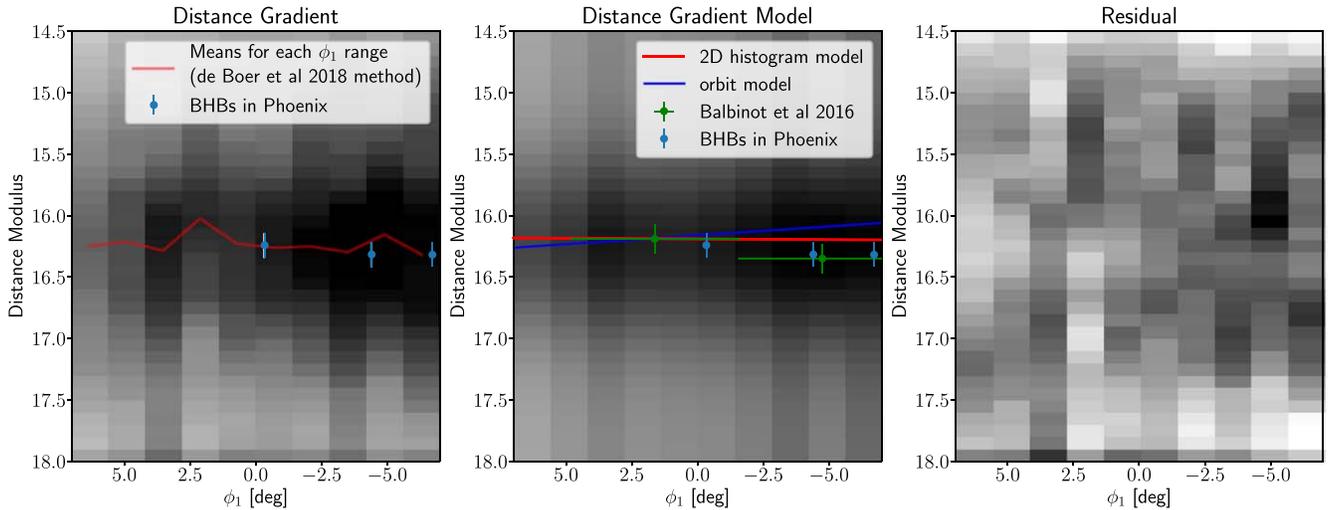

**Figure 4.** Left: a 2D histogram binning stars by $\phi_1$ and distance modulus. We bin the distance modulus using the same matched-filter selection parameters as described in Section 3 with varying $m - M$. We fit a Gaussian to the number of stars that pass the matched filter as a function of distance moduli in each $\phi_1$ bin and show the mean of each distribution with the red line. This is a method taken from de Boer et al. (2018) to simply visualize distance variations along a stream. For comparison, we also plot the distance moduli of three blue horizontal branch (BHB) stars. These are spectroscopic BHB members in Wan et al. (2020) whose distances were derived in the same way as described in Shipp et al. (2021). Middle: the histogram shows the 2D distance gradient model described in Section 4.2. We plot the line used to create that model in red. The green points show the measured distance moduli from Balbinot et al. (2016) for each half of the stream. In blue we show the distance gradient suggested by the results of the orbit model described in Section 4.3. Right: the residual between the two 2D histograms in the left and middle panels.

We model the data in Figure 4 with a two-component model consisting of a Milky Way foreground stellar population and the Phoenix stream. The foreground distribution of stars is modeled as a uniform component in distance modulus, $m - M$, for each bin of $\phi_1$. We determine the normalization of this component, $B(\phi_1)$, from the average star counts in bins that are far from the distance modulus of the Phoenix stream, using an equal number of bins on either side ($m - M < 15.5$ mag and $m - M > 17.0$ mag). The Phoenix stream is modeled with a linear-Gaussian model that depends on both $\phi_1$ and $m - M$. This model is described by three free parameters: the distance modulus at $\phi_1 = 0°$, $\mu_0$, the distance gradient, $m_\mu$, and the width of the Gaussian, $\sigma_\mu$. In each bin of $\phi_1$, an additional fixed normalization, $N_0(\phi_1)$, is taken from the linear intensity of the spline model evaluated at $\phi_1$ (Section 4.1). Thus, using $\mu = m - M$, our model for the distance of the Phoenix stream can be described by

$$\rho(\phi_1, m - M) = B(\phi_1)$$
$$+ N_0(\phi_1)\exp\left[-\frac{1}{2}\left(\frac{(m - M) - (\mu_0 + m_\mu\phi_1)}{\sigma_\mu}\right)^2\right]. \quad (2)$$

We build a binned Poisson likelihood function to compare the number of stars predicted by the model to the observed stars in each bin of $\phi_1$ and $m - M$. We sample the posterior probability distributions of the model parameters and constrain the marginalized posterior distributions of $\mu_0$ and $m_\mu$. We find best-fit values of $\mu_0 = 16.19 \pm 0.01$ (stat.) $\pm 0.1$ (sys.) mag and $m_\mu = -0.0011 \pm 0.0007$ mag deg$^{-1}$. This corresponds to a physical distance of $17.4 \pm 0.1$ (stat.) $\pm 0.8$ (sys.) kpc and a distance gradient of $-0.009 \pm 0.006$ kpc deg$^{-1}$. As with the distance modulus value reported in Section 3, the uncertainties on the distance modulus are broken into a formal statistical uncertainty and a systematic uncertainty on synthetic isochrone fitting taken from Drlica-Wagner et al. (2015). The best-fit model and residuals are shown in the middle and left panels of Figure 4, respectively.

Measurements of the Phoenix stream distance gradient are complicated by the presence of an unidentified structure overlapping the southern end of the stream at $\phi_1 \approx -7°$. This is easiest to see in Figure 2, to the lower right of the stream. As discussed in Section 3, this feature is found to have a distance modulus of $m - M = 16.2$ mag and could bias measurements of the detected gradient. For this reason, we exclude data at $\phi_1 < -7°$ from the above analysis. We also mask the Phoenix dwarf, located just below the stream at $\phi_1 \approx 5°$, in order to avoid contamination.

### 4.3. Orbit Model

We run a model of the dynamical evolution of the Phoenix stream to simulate the final debris seen in the data today and provide updated orbit parameters for the stream. To do this, we use the modified Lagrange Cloud Stripping technique from Gibbons et al. (2014), in which particles are stripped at the Lagrange points of the progenitor and evolved forward in the joint potential of the progenitor, Milky Way, and Large Magellanic Cloud (LMC). We model the Milky Way potential using the results of McMillan (2017), and we evaluate the potential using the `galpot` code (Dehnen & Binney 1998).[61] As in Shipp et al. (2021), we sample the MCMC chains from the McMillan (2017) fit and use the lowest-mass realization, with $M_{MW} = 8.3 \times 10^{11} \, M_\odot$. We include the LMC, modeled as a spherical Hernquist profile (Hernquist 1990), embedded with a Miyamoto–Nagai disk (Miyamoto & Nagai 1975). We fix the total mass of the LMC to $1.5 \times 10^{11} \, M_\odot$, motivated by the results of Erkal et al. (2019) and Shipp et al. (2021). The stream progenitor is modeled as a Plummer sphere (Plummer 1911), with a mass of $2 \times 10^4 \, M_\odot$ and a scale radius of 10 pc.

We fit the dynamical model by producing mock observations of the simulated stream and calculate the likelihood via comparison to the observed stream track, radial velocity, and proper motions. We use the stream track recovered from the stream model outlined in Section 4.1. In addition, we use $S^5$ radial velocities and Gaia EDR3 proper motions of the member stars identified by Wan et al. (2020). We incorporate the measured distance modulus of 16.19 mag (Section 4.2) as a broad Gaussian prior ($\sigma = 1.0$ mag) on the progenitor distance (at $\phi_1 = 0$ deg). This allows for an independent prediction of the distance gradient, which we show in Figure 4.

---

[61] See Table A.3 in Shipp et al. (2021) for details of this potential.

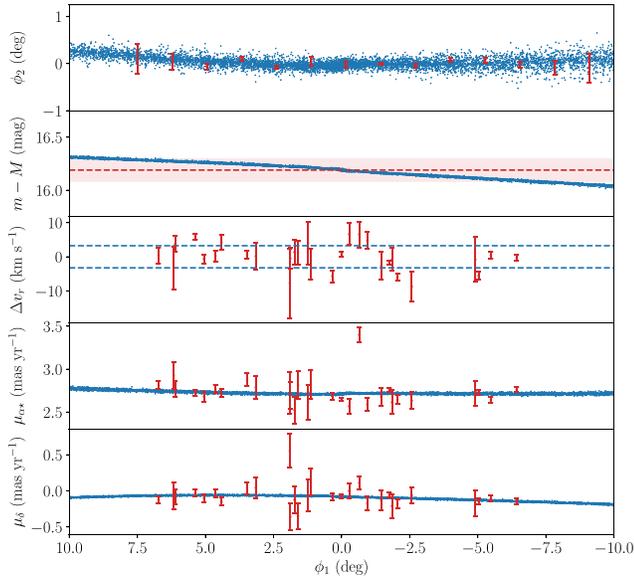

**Figure 5.** Dynamical model described in Section 4.3 fit to the Phoenix stream data. In each panel, blue represents the best-fit model, and red represents the measurements. The first panel shows the stream track measurement presented in Section 3, where the red points indicate the best-fit $\Phi_2$ value and the uncertainty at each node. In the second panel, the dotted red line and corresponding shaded area show the result of the gradient model (Section 4.2) for the distance modulus at $\phi_1 = 0°$, which was used as a prior for this model.

The best-fit orbital model is shown in Figure 5, where the blue points represent the simulated stream particles and the red points are the data included in the likelihood calculation. The orbit fit recovers values of $r_{\rm peri} = 13.0^{+0.2}_{-0.1}$ and $r_{\rm apo} = 18.6 \pm 0.1$, corresponding to an eccentricity of 0.18. This is similar to the values found by Wan et al. (2020) of $r_{\rm peri} = 12.9$, $r_{\rm apo} = 18.4$, and eccentricity = 0.18. In addition, we find a total energy $E_{\rm tot} = -0.0994^{+0.0003}_{-0.0004}$ kpc$^2$ Myr$^{-2}$ and angular momentum perpendicular to the Galactic disk $L_z = -1.65 \pm 0.01$ kpc$^2$ Myr$^{-1}$, indicating that the stream is on a prograde orbit around the Galaxy. We find the orbital pole of Phoenix to be $(\phi, \psi) = (61°8, 119°8)$, where $\phi$ and $\psi$ are the Galactic azimuthal and polar angles of the orbital pole, respectively. This result is in agreement with the one found by Riley & Strigari (2020).

The best-fit orbit model predicts a distance gradient of 0.11 kpc deg$^{-1}$. This is moderately inconsistent with our distance gradient model, which is almost flat, and the previously suggested gradient from Balbinot et al. (2016), which slopes in the opposite direction. For clarity, Balbinot et al. (2016) do not cite a distance gradient as such, but rather report two distance modulus values for the stream north and south of decl. $\delta = -50°$, concluding that a distance gradient cannot be ruled out. Interestingly, we observe that within the boundaries of our model, and notwithstanding any additional large perturbers, a gradient in the direction of that implied by Balbinot et al. (2016) is not possible.

## 5. Discussion

### 5.1. Density Variations

The track model derived in Section 4.1 demonstrates the clumpy structure of Phoenix (Figure 3). We compare the density variations in Phoenix to other narrow streams analyzed with a similar natural spline framework in Figure 6. Relative to ATLAS (Li et al. 2020) or Jet (Ferguson et al. 2021), Phoenix shows variations in linear intensity on smaller spatial scales. On the other hand, Pal 5 has large-amplitude, small-scale variations in linear intensity close to its progenitor, but only smaller-amplitude variations at larger separations. These results suggest that the large-amplitude, small-scale variations in the linear intensity that are measured in Phoenix are unusual relative to other cold streams.

We note several other traits in the structure of Phoenix. We find that the southern part of the stream ($\phi_1 < 0°$) contains, on average, a higher linear intensity than the northern section. We detect peaks at $\phi_1 \approx -1°5$, $3°0$, and $5°0$ and an underdense region (gap) at $\phi_1 \approx 1°0$. We quantify the significance of these features by using the statistical method described in Section 3.5 of Erkal et al. (2017). In contrast to the measurement of the track wiggle significance, which evaluated the significance of the wiggles over the entire stream, this process looks at each feature individually. We define $S$, the density excursion parameter, as the ratio of the density at a location $\phi_1$ to the density linearly interpolated between two background points on either side ($\phi_{1,l}$, $\phi_{1,r}$),

$$S = \frac{(\phi_{1,r} - \phi_{1,l})D(\phi_1)}{(\phi_{1,r} - \phi_1)D(\phi_{1,l}) + (\phi_1 - \phi_{1,l})D(\phi_{1,r})}. \quad (3)$$

We sample from the model posterior to evaluate the uncertainty in the $S$ statistic in order to generate a significance associated with each peak (trough) based on the probability that the $S$ statistic is greater (less) than 1. To choose the left and right points of each peak (trough), we deviate from the method used in Erkal et al. (2017) and fit a background cubic function to the linear intensity. We then find the intersection points with the model and use those for the endpoints of the window. Using this method, we recover significance values of $1.8\sigma$, $2.5\sigma$, and $1.3\sigma$ for the peaks at $\phi_1 = -1°5$, $3°0$, and $5°0$, respectively.

The large trough in the stream at $\phi_1 \approx 1°$ has a significance of $2.5\sigma$. In the region of this prominent trough, the stream density descends to its lowest point and remains there for an extended stretch of a couple degrees, in contrast to the rest of the stream density, which changes on degree scales. This difference might suggest a different origin for the gap, one that would cause a deeper and more extensive impact. It is therefore likely that this feature was caused by a more massive perturber, closer collision, or complete disruption of the progenitor (e.g., Webb & Bovy 2019), each capable of generating a large disturbance in the stream.

### 5.2. Distance Gradient

Our best-fit distance gradient, $m_\mu = -0.0011 \pm 0.0007$ mag deg$^{-1}$ ($-0.009 \pm 0.006$ kpc deg$^{-1}$), is moderately consistent with that suggested by Balbinot et al. (2016) as seen in Figure 4. It suggests almost no change in distance modulus along the stream, with a maximum difference of about 0.13 kpc

                                          

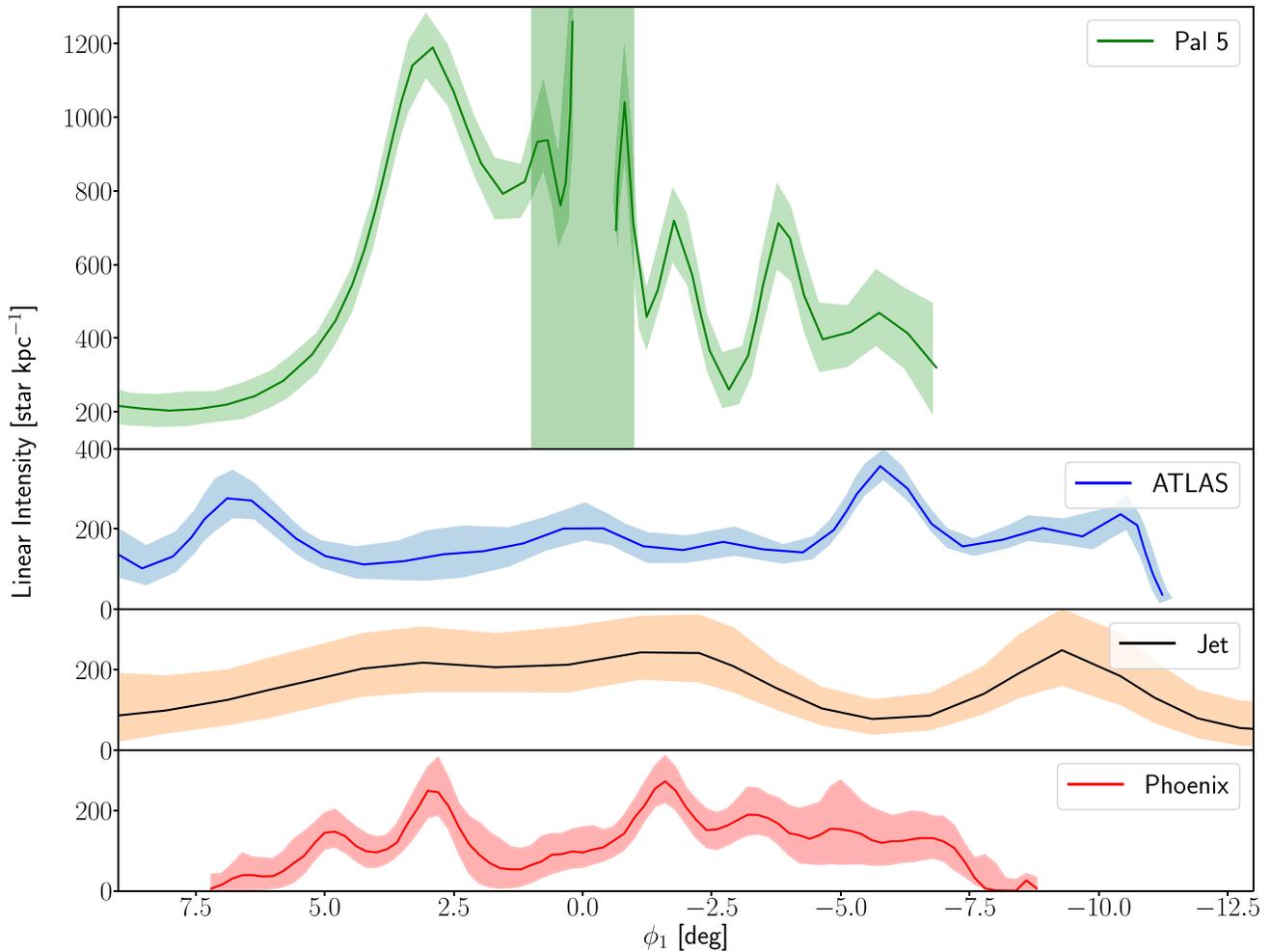

**Figure 6.** A comparison of four different streams modeled using the same parametric method as described in this work for four streams: Palomar 5 (Erkal et al. 2017), ATLAS (Li et al. 2020), Jet (Ferguson et al. 2021), and Phoenix (this work). We take the linear intensity model results directly from these three papers to make this comparison between the four streams. We note that none have been studied with data as deep as the DES Y6 data used in this analysis. The shaded areas around each stream are the minimum intervals containing 68% of the posterior probability, similar to those described and shown in Figure 3. The rectangular shaded green area corresponds to a radius of 1° around the progenitor of Pal 5, which is responsible for the density fluctuations in that region. We note that the density variations present in Phoenix are on smaller scales than those seen in ATLAS or Jet and are comparable to those within 5° of the progenitor of Pal 5.

between the northern and southern ends. We note that the larger distance gradient predicted by our orbit modeling in Section 4.3 (0.11 kpc deg$^{-1}$) is inconsistent with both our best-fit distance gradient and the gradient suggested by Balbinot et al. (2016). This suggests that our orbit model requires some additional gravitational interaction in order to produce a distance gradient that matches the observations.

This conclusion is further supported by comparing our distance gradient with the ratio of radial velocity to proper motion along the stream, as in Section 4.2 of Shipp et al. (2021). For unperturbed streams orbiting in a standard Milky Way potential, these numbers should be comparable. Within the orbit model, we do in fact find that these values are consistent along the stream. However, our measured distance gradient of $-0.009 \pm 0.006$ kpc deg$^{-1}$ is significantly offset from $v_r / \mu_{\phi_1}$, which Shipp et al. (2021) show is between 0.07 and 0.12 along the length of the stream. This difference suggests that Phoenix may have experienced a large perturbation that is not accounted for in the orbit model.

### 5.3. Stream Track Variations

Variations in the tracks of stellar streams in the $\phi_2$ direction have generally been seen on large scales, such as for Palomar 5 and ATLAS (Erkal et al. 2017; Li et al. 2020). These variations occur when tidal debris precess and nutate in an aspherical potential (e.g., Ibata et al. 2001; Helmi 2004; Johnston et al. 2005; Belokurov et al. 2014; Erkal et al. 2016). In such potentials, the stream track is expected to deviate gradually and smoothly from a great circle orbit (Erkal et al. 2016). However, such evolution is not predicted to result in small-scale features in the stream track in a galaxy such as the Milky Way (Erkal et al. 2017). This suggests that the apparent wiggles in the Phoenix stream (last panel of Figure 3) are not due to the smooth potential of the Milky Way but are instead due to smaller-scale perturbations.

One potential cause of the stream track variations is interaction with perturbers. These would likely be the same interactions that caused the density fluctuations discussed in Section 5.1. If they are, it could explain why we observe track and density deviations in the same part of the stream ($\phi_1 \sim 3°-5°$). Given Phoenix's clumpy nature, it seems possible that it has come into contact with several perturbers that affected both the density and the track. One possible outcome of these interactions is that structure could be seen in the measured radial velocities of stream members. The perturbations would have affected certain areas more than others, disturbing the motion and velocities of some stars. This would pull them away from the great circle track and manifest as wiggles, explaining the deviations in the stream track (Koposov et al. 2019). Lastly, we note the possibility of a perturbation caused by the progenitor. The progenitor is known to cause stream track deviations in other streams such as Palomar 5 (Erkal et al. 2017; Bonaca et al. 2020). Recently, Li et al. (2020) showed that wiggles on the scale of tenths of degrees could be caused by a progenitor. That may explain one of these wiggles observed in Phoenix.

### 5.4. Origins of Density Variations

Several mechanisms have been proposed to create peaks and gaps in stellar streams similar to those that we observe in Phoenix. Determining the origin of each inhomogeneity can help us determine the ways in which various structures in the Milky Way halo interact with stellar streams.

#### 5.4.1. Stream Formation

Density variations can be introduced in a variety of ways, including through the process of stream formation. As stars escape from the progenitor to create tidal features, their epicyclic motion can slow as they get farther away. This tends to happen at a similar location for many stars because of the similarity in initial positions and velocities, leading to a clump (Küpper et al. 2008). In theory, for each orbital passage, this could create an overdensity at some distance from the progenitor. This distance depends on the stream's orbit, the progenitor mass, and the strength of the tidal field (Küpper et al. 2010). This leads to the possibility that periodic clumps along the stream are formed in this way.

Density variations can also be introduced during formation if the Phoenix stream originates from a globular cluster that accretes within a parent satellite galaxy. Malhan et al. (2020) suggest that streams formed from globular clusters in accreted satellite galaxies may be perturbed by the parent satellite. Because the stream and its parent satellite are on similar orbits, there is a longer time where the two structures are in close proximity to one another and the gravity from the parent subhalo can affect the uniformity of the stream. In addition to causing gaps in the stream, a similar process could lead to narrow spikes in the stream density because as the subhalo orbits the Milky Way, variations in its mass-loss rate cause it to episodically deposit increased amounts of globular cluster stellar debris (Malhan et al. 2020). The narrow width, low velocity dispersion, and low metallicity of the Phoenix stream may suggest that its progenitor was a globular cluster formed and accreted as part of a low-mass galaxy (e.g., Balbinot et al. 2016).

#### 5.4.2. Milky Way Structure and Halo Perturbers

It is also possible for streams to be perturbed by baryonic structures in the Milky Way. This includes the Milky Way bar, which can cause different stars along the stream to reach pericenter at different times. A rotating bar causes these stars to experience unequal torques, leading to a redistribution of energy along the stream (Hattori et al. 2016). This in turn leads to different orbital periods along the stream, potentially resulting in ever-widening gaps in the stream intensity (Erkal et al. 2017; Pearson et al. 2017).

Other known perturbers are satellite galaxies such as the Large and Small Magellanic Clouds (LMC and SMC) and Sagittarius (Garavito-Camargo et al. 2019; Petersen & Peñarrubia 2020, 2021). In Phoenix's case, Shipp et al. (2021) determined that the stream should have experienced less significant perturbations from the LMC than most other streams owing to its large distance of closest approach and high relative velocity. In addition to direct perturbations, the LMC creates a reflex motion in the Milky Way that can affect streams' orbits and has been shown to do so for Sagittarius (Gómez et al. 2015; Vasiliev et al. 2021). We note that the models used in this work (Section 4.3) do account for the reflex motion of the Milky Way. Milky Way globular clusters, although smaller, can also exert gravitational influence on streams. In fact, because of their smaller size, they are more likely than larger objects like the LMC or SMC to induce smaller-scale variations of the kind seen in Phoenix. Lastly, giant molecular clouds (GMCs) have been shown to affect stellar streams (Amorisco et al. 2016). N-body simulations have shown that GMC disks can cause gaps and clumps in nearby streams. This analysis is performed on Pal 5 and GD-1 by Amorisco et al. (2016), who conclude that these GMCs could cause disturbances in cold thin streams similar to those created by dark matter subhalo flybys. However, Li et al. (2020) look at the effect of GMCs on ATLAS, which is prograde with a pericenter of ~13 kpc, and concluded that they created a maximum wiggle of 0°.04. Therefore, we find it unlikely that GMCs caused any of the features in the Phoenix stream.

Dark matter subhalos in the Milky Way are also among the possible sources of perturbation to the density of a stellar stream. When simulating the effect of a subhalo on idealized streams, it is clear that such structures can cause both large and small variations in the stream density (Ibata et al. 2002; Johnston et al. 2002). Carlberg (2020) finds that modeling the interaction between a stream and a full population of dark matter subhalos, which includes halos of mass below $4 \times 10^8$ $M_\odot$, causes a significant increase in smaller-scale density fluctuations when compared to models that do not include subhalos of mass $<4 \times 10^8 \, M_\odot$.

Streams can be used to inform our understanding of the population of dark matter subhalos around the Milky Way (e.g., Banik et al. 2021b). For instance, Banik et al. (2021a) show that the CDM paradigm predicts a dark matter subhalo population that sufficiently explains perturbations in the GD-1 stream. Through the detection of such perturbations, it is then possible to set novel constraints on dark matter particle physics scenarios (Banik et al. 2021a). More specifically, we can reliably determine a subhalo's mass and size based on the gap it creates in a stellar stream (Erkal & Belokurov 2015b). Examining the density variations also makes it possible to track the location of subhalos (Bonaca et al. 2019). The continued study of density variations along stellar streams such as Phoenix may provide crucial insights into the properties of dark matter subhalos (Montanari & García-Bellido 2022).

The mechanisms discussed in Section 5.4 can generically introduce structure into any stellar stream. Thus, they do not inherently explain the differences in the overall clumpiness of Phoenix relative to other streams. If the extra structure in Phoenix is caused by one of these mechanisms, it is necessary to explain what unique orbital or evolutionary properties of Phoenix have resulted in smaller-scale density variations than are observed in other streams. Answering questions about the formation of density variations in streams will most likely require a concerted effort to assemble a catalog of similarly modeled streams. Such an undertaking will be enabled by deeper data covering more of the sky and allowing similar analyses to be performed on fainter streams (e.g., from the Rubin Observatory; Ivezić et al. 2019). Such work, combined with the devoted modeling of individual stream perturbation events (e.g., Bonaca et al. 2019), could provide valuable insight into the formation and evolution of stellar streams, as well as the nature of dark matter.

## 6. Conclusions

In this paper, we use the 6 yr of data from DES to conduct a deeper analysis of the Phoenix stellar stream. We apply a nonparametric stream model using natural splines to determine stream properties such as the stream track, width, and linear density. Our model makes few a priori assumptions about the structure of the Phoenix stream, and therefore the model complexity comes directly from the data. This model prefers a clumpy stream density distribution throughout the entire length of the stream. These clumps and gaps vary in angular size, as well as in their maximum and minimum linear intensities. We also observe wiggles in the stream track, which we determine to have a significance of ~1.2σ. Such small-scale wiggles should be a rare occurrence in stellar streams in aspherical host potentials (Erkal et al. 2016). We also measure the distance and distance gradient of the Phoenix stream and find that it has a very small gradient, $-0.009 \pm 0.006$ kpc deg$^{-1}$. Our measured distance gradient is in moderate disagreement with an orbital model built on the track and mean distance of the Phoenix stream and the proper motion and radial velocities of stream members, suggesting that the orbital model does not incorporate some additional gravitational interaction Phoenix had, which would flatten its gradient.

We consider the potential causes of the variations we find in the density and track of the Phoenix stream. In particular, we discuss the possibility that the stream may have interacted with baryonic perturbers or dark matter subhalos that could have left behind the clumps, gaps, and wiggles we observed. We also consider the possibility that Phoenix may have originated from a globular cluster that was accreted as part of a dwarf galaxy system. The relatively large number of clumps and fluctuations relative to other streams could indicate that disturbances from the parent satellite may have played a role in producing the complex stream structure we observe today.

The purpose of this paper is to lay the groundwork for future analysis on the formation and evolution of the Phoenix stellar stream. By providing a more detailed analysis of the stream using deeper data, our results will enable others to run more advanced models to simulate the clumps and wiggles found here. Such work would advance our understanding of the mechanisms for introducing variations into stellar streams and form the groundwork for performing similar analyses on other streams.

K.T. acknowledges support from the Provost Scholars Program at the University of Chicago. P.S.F. acknowledges support from the Visiting Scholars Award Program of the Universities Research Association.

Funding for the DES Projects has been provided by the U.S. Department of Energy, the U.S. National Science Foundation, the Ministry of Science and Education of Spain, the Science and Technology Facilities Council of the United Kingdom, the Higher Education Funding Council for England, the National Center for Supercomputing Applications at the University of Illinois at Urbana-Champaign, the Kavli Institute of Cosmological Physics at the University of Chicago, the Center for Cosmology and Astro-Particle Physics at The Ohio State University, the Mitchell Institute for Fundamental Physics and Astronomy at Texas A&M University, Financiadora de Estudos e Projetos, Fundação Carlos Chagas Filho de Amparo à Pesquisa do Estado do Rio de Janeiro, Conselho Nacional de Desenvolvimento Científico e Tecnológico and the Ministério da Ciência, Tecnologia e Inovação, the Deutsche Forschungsgemeinschaft, and the Collaborating Institutions in the Dark Energy Survey.

The Collaborating Institutions are Argonne National Laboratory, the University of California at Santa Cruz, the University of Cambridge, Centro de Investigaciones Energéticas, Medioambientales y Tecnológicas-Madrid, the University of Chicago, University College London, the DES-Brazil Consortium, the University of Edinburgh, the Eidgenössische Technische Hochschule (ETH) Zürich, Fermi National Accelerator Laboratory, the University of Illinois at Urbana-Champaign, the Institut de Ciències de l'Espai (IEEC/CSIC), the Institut de Física d'Altes Energies, Lawrence Berkeley National Laboratory, the Ludwig-Maximilians Universität München and the associated Excellence Cluster Universe, the University of Michigan, the National Optical Astronomy Observatory, the University of Nottingham, The Ohio State University, the University of Pennsylvania, the University of Portsmouth, SLAC National Accelerator Laboratory, Stanford University, the University of Sussex, Texas A&M University, and the OzDES Membership Consortium.

Based in part on observations at Cerro Tololo Inter-American Observatory, National Optical Astronomy Observatory, which is operated by the Association of Universities for Research in Astronomy (AURA) under a cooperative agreement with the National Science Foundation.

The DES data management system is supported by the National Science Foundation under grant Nos. AST-1138766 and AST-1536171. The DES participants from Spanish institutions are partially supported by MINECO under grants AYA2015-71825, ESP2015-66861, FPA2015-68048, SEV-2016-0588, SEV-2016-0597, and MDM-2015-0509, some of which include ERDF funds from the European Union. IFAE is partially funded by the CERCA program of the Generalitat de Catalunya. Research leading to these results has received funding from the European Research Council under the European Union's Seventh Framework Program (FP7/2007–2013) including ERC grant agreements 240672, 291329, and 306478. We acknowledge support from the Australian Research Council Centre of Excellence for All-sky Astrophysics (CAASTRO), through project No. CE110001020, and the Brazilian Instituto Nacional de Ciência e Tecnologia (INCT) e-Universe (CNPq grant 465376/2014-2).

This manuscript has been authored by Fermi Research Alliance, LLC, under contract No. DE-AC02-07CH11359 with the U.S. Department of Energy, Office of Science, Office of High Energy Physics. The United States Government retains and the publisher, by accepting the article for publication, acknowledges that the United States Government retains a nonexclusive, paid-up, irrevocable, worldwide license to publish or reproduce the published form of this manuscript, or allow others to do so, for United States Government purposes.

This work has made use of data from the European Space Agency (ESA) mission Gaia (https://www.cosmos.esa.int/gaia), processed by the Gaia Data Processing and Analysis Consortium (DPAC, https://www.cosmos.esa.int/web/gaia/dpac/consortium). Funding for the DPAC has been provided by national institutions, in particular the institutions participating in the Gaia Multilateral Agreement.

Based on data acquired at the Anglo-Australian Telescope. We acknowledge the traditional owners of the land on which the AAT stands, the Gamilaroi people, and pay our respects to elders past and present.

## Appendix
## Verifying the Model with ATLAS

In order to verify the proper functioning of our slightly updated model from versions used for previous streams, we attempt to replicate the results from Li et al. (2020) for the ATLAS stream. We use the same 3 yr DES data used in that paper and run the analysis we intend to run on Phoenix. We first determine the optimal number of nodes for each of the six parameters described in Section 4, before running a maximum log-likelihood optimization for the best-fit curves of each parameter. Due to the complexity of the six parameters, the limited run time of the node optimization, and the different cross-validation subsets used in our analysis versus the published one, we do not recover the exact same number of nodes for each parameter. However, we find excellent agreement (green and blue curves in Figure A1) in the curves for the intensity, track, and width with the published results from Li et al. (2020). The ability to replicate previous results builds confidence in our ability to characterize Phoenix in a manner similar to that previously applied to other streams.

Before applying the model to Phoenix, we also verify that switching to the new and deeper DES Y6 data set does not have any artificial impact on the stream features by running the same optimization on the ATLAS stream but with the new data. We again recover slightly different numbers of nodes but an outcome that agrees with the previously published results and our reanalysis of the Y3 data as seen in Figure A1.

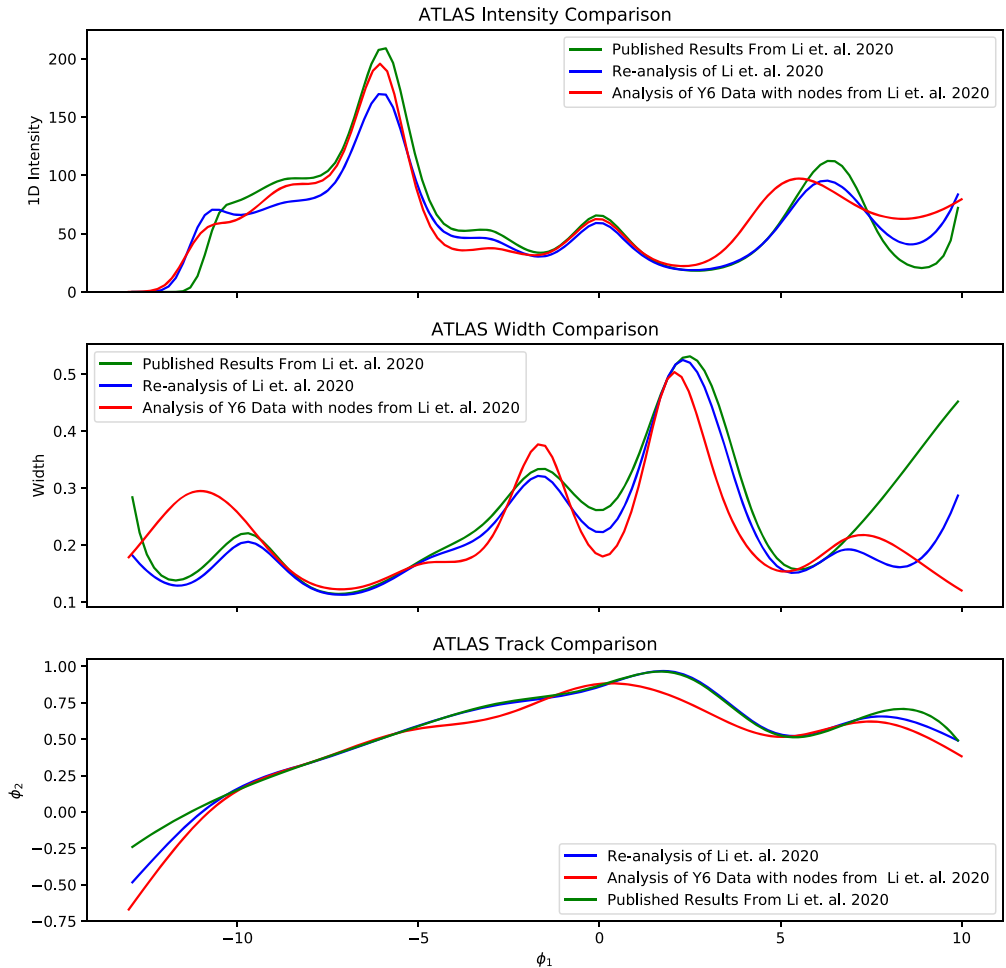

**Figure A1.** We test that our model works properly by replicating the results from Li et al. (2020). We show a comparison between the 1D intensities, widths, and tracks in the three panels. The green curve shows the published results from Li et al. (2020). The blue curve shows our reanalysis using the same Y3 data as that paper used. The red curve shows an analysis using the deeper Y6 data. We clearly see a good agreement for all three curves. This gives us confidence that our model works properly and that the new data do not have any artificial impacts on the stream features.

## ORCID iDs

K. Tavangar  https://orcid.org/0000-0001-6584-6144
P. Ferguson  https://orcid.org/0000-0001-6957-1627
N. Shipp  https://orcid.org/0000-0003-2497-091X
A. Drlica-Wagner  https://orcid.org/0000-0001-8251-933X
S. Koposov  https://orcid.org/0000-0003-2644-135X
D. Erkal  https://orcid.org/0000-0002-8448-5505
E. Balbinot  https://orcid.org/0000-0002-1322-3153
J. García-Bellido  https://orcid.org/0000-0002-9370-8360
K. Kuehn  https://orcid.org/0000-0003-0120-0808
G. F. Lewis  https://orcid.org/0000-0003-3081-9319
T. S. Li  https://orcid.org/0000-0002-9110-6163
S. Mau  https://orcid.org/0000-0003-3519-4004
A. B. Pace  https://orcid.org/0000-0002-6021-8760
A. H. Riley  https://orcid.org/0000-0001-5805-5766
M. Aguena  https://orcid.org/0000-0001-5679-6747
S. Allam  https://orcid.org/0000-0002-7069-7857
J. Annis  https://orcid.org/0000-0002-0609-3987
E. Bertin  https://orcid.org/0000-0002-3602-3664
D. Brooks  https://orcid.org/0000-0002-8458-5047
D. L. Burke  https://orcid.org/0000-0003-1866-1950
A. Carnero Rosell  https://orcid.org/0000-0003-3044-5150
M. Carrasco Kind  https://orcid.org/0000-0002-4802-3194
J. Carretero  https://orcid.org/0000-0002-3130-0204
M. Costanzi  https://orcid.org/0000-0001-8158-1449
J. De Vicente  https://orcid.org/0000-0001-8318-6813
H. T. Diehl  https://orcid.org/0000-0002-8357-7467
B. Flaugher  https://orcid.org/0000-0002-2367-5049
J. Frieman  https://orcid.org/0000-0003-4079-3263
E. Gaztanaga  https://orcid.org/0000-0001-9632-0815
D. W. Gerdes  https://orcid.org/0000-0002-6942-2736
D. Gruen  https://orcid.org/0000-0003-3270-7644
J. Gschwend  https://orcid.org/0000-0003-3023-8362
G. Gutierrez  https://orcid.org/0000-0003-0825-0517
K. Honscheid  https://orcid.org/0000-0002-6550-2023
D. J. James  https://orcid.org/0000-0001-5160-4486
N. Kuropatkin  https://orcid.org/0000-0003-2511-0946
M. A. G. Maia  https://orcid.org/0000-0001-9856-9307
J. L. Marshall  https://orcid.org/0000-0003-0710-9474
F. Menanteau  https://orcid.org/0000-0002-1372-2534
R. Miquel  https://orcid.org/0000-0002-6610-4836
R. L. C. Ogando  https://orcid.org/0000-0003-2120-1154
A. Palmese  https://orcid.org/0000-0002-6011-0530
F. Paz-Chinchón  https://orcid.org/0000-0003-1339-2683
A. Pieres  https://orcid.org/0000-0001-9186-6042
A. A. Plazas Malagón  https://orcid.org/0000-0002-2598-0514
E. Sanchez  https://orcid.org/0000-0002-9646-8198
I. Sevilla-Noarbe  https://orcid.org/0000-0002-1831-1953
M. Smith  https://orcid.org/0000-0002-3321-1432
E. Suchyta  https://orcid.org/0000-0002-7047-9358
G. Tarle  https://orcid.org/0000-0003-1704-0781
C. To  https://orcid.org/0000-0001-7836-2261
A. R. Walker  https://orcid.org/0000-0002-7123-8943